\documentclass[
  aps,
  prl,
  reprint,
  superscriptaddress,
  amsmath,amssymb,
]{revtex4-2}
\usepackage{graphicx}
\usepackage{bm}
\usepackage{siunitx}
\usepackage{xcolor}
\usepackage{hyperref}
\usepackage{natbib}

\usepackage[normalem]{ulem}

\newcommand{\vprfl}{\tilde{V}_\mathrm{RF}^\mathrm{L}}
\newcommand{\diff}{\mathrm{d}}
\newcommand{\teff}{t_\mathrm{eff}}
\newcommand{\vbs}{V_\mathrm{BS}}
\newcommand{\vbd}{V_\mathrm{BD}}
\newcommand{\vbt}{V_\mathrm{BT}}
\newcommand{\vbb}{V_\mathrm{BB}}
\newcommand{\vpl}{V_\mathrm{L}}
\newcommand{\vpr}{V_\mathrm{R}}
\newcommand{\vrf}{V_\mathrm{RF}}

\DeclareSIUnit{\belmilliwatt}{Bm}
\DeclareSIUnit{\dBm}{\deci\belmilliwatt}
\DeclareSIUnit{\arbunit}{arb.\,units}

\newcommand{\subref}[2]{\hyperref[#1]{\ref*{#1}#2}}
\def\problem#1{\if\problemmarker1\textcolor{red}{#1}\else\errmessage{#1}#1\fi}
\let\problemmarker=1

\begin{document}
\title{Flux-Tunable Hybridization in a Double Quantum Dot Interferometer}

\author{Christian G. Prosko}
\affiliation{QuTech and Kavli Institute of Nanoscience, Delft University of Technology, 2600 GA Delft, The Netherlands}

\author{Ivan Kulesh}
\affiliation{QuTech and Kavli Institute of Nanoscience, Delft University of Technology, 2600 GA Delft, The Netherlands}

\author{Michael Chan}
\affiliation{QuTech and Kavli Institute of Nanoscience, Delft University of Technology, 2600 GA Delft, The Netherlands}

\author{Lin Han}
\affiliation{QuTech and Kavli Institute of Nanoscience, Delft University of Technology, 2600 GA Delft, The Netherlands}

\author{Di Xiao}
\affiliation{Department of Physics and Astronomy, Purdue University, West Lafayette, Indiana 47907, USA}

\author{Candice Thomas}
\affiliation{Department of Physics and Astronomy, Purdue University, West Lafayette, Indiana 47907, USA}

\author{Michael J. Manfra}
\affiliation{Department of Physics and Astronomy, Purdue University, West Lafayette, Indiana 47907, USA}
\affiliation{School of Materials Engineering, Purdue University, West Lafayette, Indiana 47907, USA}
\affiliation{Elmore School of Electrical and Computer Engineering,
  Purdue University, West Lafayette, Indiana 47907, USA}

\author{Srijit Goswami}
\affiliation{QuTech and Kavli Institute of Nanoscience, Delft University of Technology, 2600 GA Delft, The Netherlands}

\author{Filip K. Malinowski}
\affiliation{QuTech and Kavli Institute of Nanoscience, Delft University of Technology, 2600 GA Delft, The Netherlands}

\date{\today}

\begin{abstract}
  A single electron shared between two levels threaded by a magnetic flux is an irreducibly simple quantum system in which interference is predicted to occur.
  We demonstrate tuning of the tunnel coupling between two such electronic levels with flux, implemented in a loop comprising two quantum dots.
  Using radio-frequency reflectometry of the dots' gate electrodes we extract the inter-dot coupling, which exhibits oscillations with a periodicity of one flux quantum.
  In different tunneling regimes we benchmark the oscillations' contrast, and find that their amplitude varies with the levels involved, while tunneling is generically not suppressed at oscillation minima.
  These results establish the feasibility and limitations of parity readout of qubits with tunnel couplings tuned by flux.
\end{abstract}

\maketitle


Phase interference between electron trajectories manifests in commonly observed phenomena such as the Aharonov-Bohm (AB) effect and weak localization \cite{Ihn_2009}.
Tunneling effects on interference have also been explored by embedding quantum dots (QDs) into AB interferometers \cite{Yacoby_1995,Schuster_1997,Holleitner_2001,Sigrist_2004,Avinun_Kalish_2005,Hatano_2011,Edlbauer_2017,Borsoi_2020}, and the interference of tunneling paths has been probed in other systems \cite{Noguchi_2014,Parto_2019,Venkatraman_2022_ArXiv}.
To date, however, the phase modulation of tunnel couplings between discrete fermionic levels has not been directly observed.
Magnetic fields in particular impart a phase on tunnel couplings between QDs.
Destructive interference of this phase may then suppress the effective coupling between them when tunneling involves multiple paths.
Importantly, symmetrically-tuned tunnel barriers in QD devices threaded by magnetic flux are necessary to maximize the readout sensitivity of measurement-based topological qubits formed in hybrid nanowires or Kitaev chains \cite{Leijnse_2012,Sz_chenyi_2020,Smith_2020,Grimsmo_2019} and for certain tests of Majorana fusion rules \cite{Karzig_2017,Liu_2022_ArXiv}.
Additionally, it has been proposed that new types of semiconducting qubits could exploit flux-tunable couplings to implement gate operations and noise-protected readout schemes \cite{Weichselbaum_2004,Wang_2006,Shim_2022}.
These interference effects are also unavoidable for arrays of QDs proposed for quantum computation and simulation \cite{Li_2018,Borsoi_2022_ArXiv}, since distant QDs may be coherently coupled by multiple paths \cite{Braakman_2013,Dehollain_2020}.
Hence, control of this phase is crucial.

We probe electronic quantum interference in an irreducibly simple case of tunneling between two electronic levels in a loop formed by two QDs.
Radio-frequency (RF) gate reflectometry is sensitive to tunnel couplings between QDs \cite{Cottet_2011,Frey_2012,Colless_2013,Urdampilleta_2015,Mizuta_2017,Talbo_2018,Esterli_2019,de_Jong_2019,Ezzouch_2021,de_Jong_2021}, so we employ it to quantify the inter-dot coupling as a function of magnetic flux, and show that it oscillates with $h/e$ periodicity.
The specific levels involved in tunneling affect the oscillation amplitude and mean tunnel coupling.
Additionally, we observe that the measured signal's dependence on tunnel coupling is nonlinear and not one-to-one, rendering changes in tunnel couplings less resolvable in stronger tunneling regimes relative to weak tunneling regimes.
This is consistent with Landau-Zener transitions (LZTs) and thermal excitations sharply suppressing the signal for very small tunnel couplings \cite{de_Jong_2019}.

\begin{figure}[ht!]
  \centering
  \includegraphics{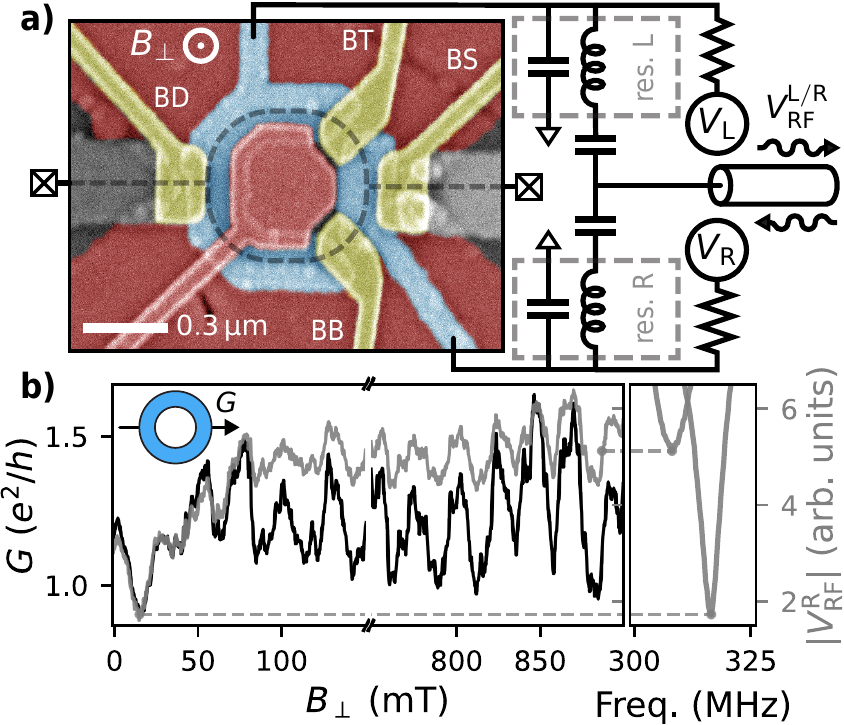}
  \caption{
    {\bf Experimental design and AB oscillations.}
    \textbf{a)} False-color electron micrograph of a nominally equivalent device to the one measured, and a schematic of the resonator circuit.
    The device may be tuned by depletion (red) and barrier (yellow, labeled) gate electrodes into an open AB loop, a ring-shaped QD, or a double quantum dot (DQD) with QDL and QDR chemical potentials tuned by plunger gate voltages $\vpl$ and $\vpr$ (blue).
    Outer and inner depletion gates have \SI{-2}{\volt} and \SI{-3}{\volt} applied respectively to form a conducting loop, illustrated by a dashed line.
    \textbf{b)} AB oscillations in the open loop configuration depicted in the inset. 
    Measurements are at zero bias voltage of 4-terminal lock-in conductance (black) and of the absolute reflected signal $\vert \vrf^{R}\vert$ (gray) from the resonator coupled to the $\vpr$ electrode at the field-dependent resonance dip.
    On the right, example frequency sweeps from which the minimizing RF signal is calculated are shown.
    $h/e$ and $h/2e$-periodic oscillations are visible in both the conductance and in the depth of the resonance.
  }
  \label{fig:setup}
\end{figure}

To fabricate a device capable of being tuned into a ring-shaped DQD, we utilize a ternary $\mathrm{InSb}_{0.86}\mathrm{As}_{0.14}$ two-dimensional electron gas (2DEG) \cite{Moehle_2021}.
The device (Fig.~\ref{fig:setup}{a}) consists of three Ti/Pd gate layers each separated by \SI{20}{\nano\meter} of $\mathrm{Al}_2\mathrm{O}_3$ dielectric patterned on the 2DEG.
Charge is confined to an annular ring geometry by depleting carriers below the outer and inner depletion gates, since the 2DEG conducts without applied voltages.
Barrier gate voltages $\vbs$, $\vbd$, $\vbt$, and $\vbb$ define a large curved QD and a smaller QD (denoted QDL and QDR, respectively), while plunger gate voltages $\vpl$ and $\vpr$ control their chemical potentials.
Two additional unlabeled gates control charge density in the exposed 2DEG between the QDs and Al contacts (Suppl. Sec.~SI).

Both plunger gates controlling QDL and QDR are bonded to resonators formed by NbTiN spiral inductors with \SI{420}{\nano\henry} and \SI{730}{\nano\henry} inductance and their parasitic capacitances, leading to resonance frequencies of roughly \SI{400}{\mega\hertz} and \SI{315}{\mega\hertz}, respectively \cite{Hornibrook_2014}.
Tunneling in the device results in frequency shifts $\Delta f_0^{\mathrm{L/R}}$ of the resonators labeled L and R due to parametric capacitance, while excess dissipation lowers their internal quality factors.
Both properties affect their reflection coefficients, which we measure using standard reflectometry techniques \cite{Vigneau_2023}.
Low-power signals reflected from the device are amplified by a high-electron-mobility transistor at \SI{4}{\kelvin} and measured with a vector network analyzer or ultra-high-frequency lock-in amplifier to produce demodulated complex signals $V_{\mathrm{RF}}^\mathrm{L/R}$, see Fig.~\subref{fig:setup}{a}.
Measurements are conducted at the approximately \SI{20}{\milli\kelvin} base temperature of a dilution refrigerator.


We begin by verifying the electron phase coherence in our device manifested by the AB effect \cite{Ihn_2009}.
To form an open loop without QDs, we set all accumulation, plunger, and barrier gates to positive voltages to remove potential barriers.
Fig.~\subref{fig:setup}{b} presents the four-terminal conductance $G$ and response of resonator R as a function of the out-of-plane field $B_\perp$.
Resonator R is sensitive to dissipative transport in the loop despite being capacitively coupled, manifesting as a reduction of the resonator's quality factor.
Matching AB ocillations and higher harmonics are prominent in both $G$ and the depth of the minimum in resonator R's reflection coefficient on resonance \cite{Pieper_1994}.
We observe a varying $\phi_0\equiv h/e$ flux periodicity consistent with the bounds expected based on the lithographically defined \SI{180}{\nano\meter} and \SI{374}{\nano\meter} inner and outer radii of the loop.
Additionally, the open loop can be transformed into a single QD by lowering $\vbs$ and $\vbd$ to form tunnel barriers, exhibiting a finite level spacing and $h/e$-periodic oscillations of the addition energy (Suppl. Sec.~SII) \cite{Fuhrer_2001}.

\begin{figure}[t]
    \centering
    \includegraphics{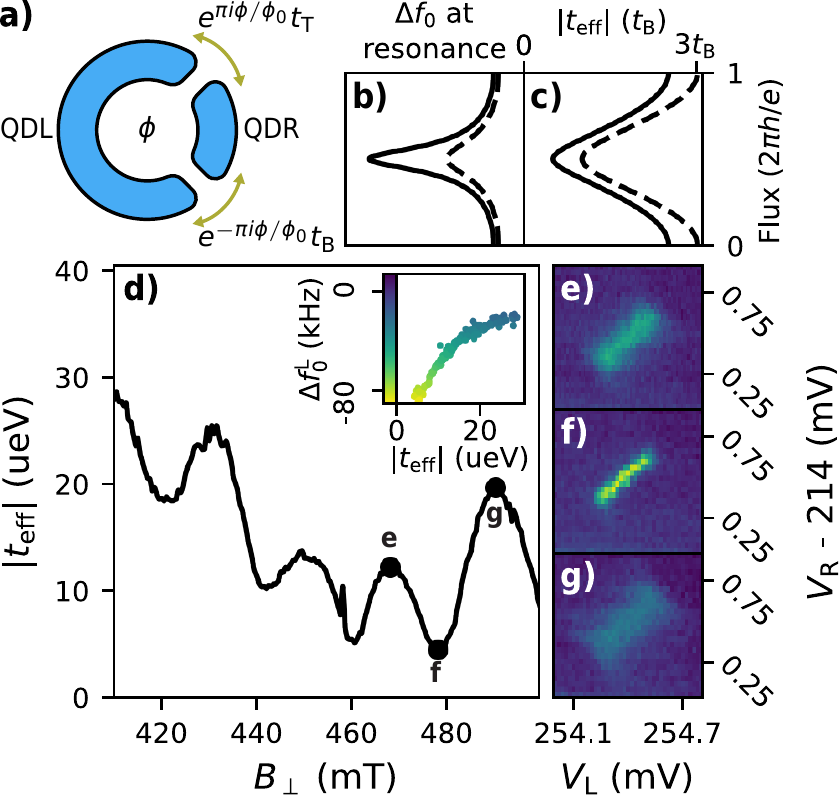}
    \caption{
      {\bf Tuning DQD hybridization with flux.}
      \textbf{a)} Diagram of a DQD ring threaded by a magnetic flux $\phi(B_\perp)$.
      \textbf{b,c)} Schematic mapping of $\vert\teff\vert$ as a function of magnetic flux $\phi$ \textbf{(c)} into a final resonator frequency shift $\Delta f_0(\phi)$ at charge resonance \textbf{(b)}, shown for $t_\mathrm{T}=1.5t_\mathrm{B}$ (solid) and $2t_\mathrm{B}$ (dashed).
      For sizable $\vert\teff\vert$ the frequency shift is $\propto 1/\vert\teff\vert$ \cite{Mizuta_2017,Esterli_2019}.
      \textbf{d)} Fit $\vert\teff\vert$ values from the frequency  response of resonator L as a function of $B_\perp$ for a single inter-dot transition.
      The tunnel coupling oscillates periodically with varying contrast and amplitude.
      The inset defines the charge stability diagram (CSD) color scale and plots the approximately $\propto 1/\vert\teff\vert$ correspondence between the fit $\vert\teff\vert$ and maximum observed $\Delta f_0^\mathrm{L}$ for each $B_\perp$ in \textbf{(d)}.
      \textbf{e,f,g)} Select CSDs at the $B_\perp$ values labeled in \textbf{(d)} showing the lineshape of $\Delta f_0^\mathrm{L}$ across the inter-dot transitions for different tunnel couplings.
    }
    \label{fig:freqfits}
\end{figure}

Having established phase coherence of the InSbAs loop, we next consider the case of a loop comprising two quantum dots threaded by a magnetic flux, illustrated in Fig.~\subref{fig:freqfits}{a}.
For this system, we expect magnetic flux to tune the effective inter-dot tunnel coupling.
Assuming that at a given inter-dot charge transition each QD is described by a single fermionic level, the DQD can be represented as a two-level system with an effective coupling matrix element $\teff \equiv t_\mathrm{T} + t_\mathrm{B}$.
Here, we define $t_\mathrm{T}$ and $t_\mathrm{B}$ as the inter-dot coupling due to the top and bottom arms, respectively.
Under the Peierls substitution, a magnetic flux $\phi(B_\perp)$ imparts a phase on each coupling \cite{Hofstadter_1976}.
Using an appropriate choice of gauge, we then have
\begin{equation}\label{eq:teff}
  \vert \teff\vert = \sqrt{\vert t_\mathrm{T}\vert^2 + \vert t_\mathrm{B}\vert^2 + 2\vert t_\mathrm{T}t_\mathrm{B}\vert \cos{(2\pi\phi/\phi_0)}},
\end{equation}
assuming $t_\mathrm{T}$ and $t_\mathrm{B}$ had equal phases at zero field.

Via quantum capacitance, $\teff(\phi)$ imparts a frequency shift on QDL's gate resonator with a maximal value in the ground state $\propto 1/\vert \teff\vert$.
Hence, we expect the frequency shift to oscillate periodically with $\phi$.
In Figs.~\subref{fig:freqfits}{b-c}, we plot the expected maximal frequency shift according to a model including thermal excitations \cite{Mizuta_2017,Esterli_2019}.

Experimentally, we realize this system as a loop-shaped DQD with chemical potentials tuned by voltages $\vpl$ and $\vpr$.
To focus on interdot transitions where the signal contains information about the inter-dot tunnel coupling $\teff$, we lower $\vbs$ and $\vbd$ until tunneling rates to the leads are undetectably small.
Meanwhile, we form the DQD by lowering $\vbt$ and $\vbb$ into a regime of moderate tunneling, such that inter-dot transitions exhibit a substantial quantum capacitance signal.
The barriers are tuned to be approximately equal based on DC current measurements (Suppl. Sec.~SIII).
Coulomb diamond measurements demonstrate a varying but finite level spacing above \SI{70}{\micro\electronvolt} in both QDs (Suppl. Sec.~SIV), such that the DQD is well-described by two coupled fermionic levels.
Maintaining a finite excitation energy on both QDs despite their large lithographic size is achievable due to the low effective mass of roughly $0.016m_\mathrm{e}$ in the 2DEG \cite{Moehle_2021}, favoring confinement.

Selecting a single inter-dot transition in this regime, we measure gate and frequency dependent traces of resonator L's response $\vrf^\mathrm{L}$ as a function of $B_\perp$, aiming to extract $\vert\teff\vert$.
At each point in gate space, we fit the results to an asymmetric resonator model to extract the resonance frequency shift $\Delta f_0^\mathrm{L}$ \cite{Khalil_2012,Probst_2015,Guan_2020}.
We fit the $\vpl$ dependence of $\Delta f_0^\mathrm{L}$ to a thermal quantum capacitance model to extract $\vert\teff\vert$, where the lever arm (0.18) and electron temperature (\SI{71}{\milli\kelvin}) parameters are optimized simultaneously for all field values to produce the minimal fit error (Suppl. Sec.~SV) \cite{Mizuta_2017,Esterli_2019}.
Subsequently they are fixed, with the only other free parameters being offsets in $\vpl$ and $\Delta f_0^\mathrm{L}$.

The resulting $\vert\teff\vert$ values are plotted in Fig.~\subref{fig:freqfits}{d}, where oscillations in $\vert\teff\vert$ are clearly visible.
In Figs.~\subref{fig:freqfits}{e-g}, we show examples of frequency shifts of resonator L for some $B_\perp$ values, where we see that for smaller tunnel couplings the transition appears more narrow but with a stronger frequency shift.
Notably, the tunnel coupling in general does not reach zero at its minima, suggesting that $t_\mathrm{T}$ and $t_\mathrm{B}$ are not precisely equal, as exemplified in Fig.~\subref{fig:freqfits}{c}.
The average value of $\vert\teff\vert$ between oscillations also varies unpredictably, indicating that the involved states' wavefunctions vary across multiple flux periods.
Nonetheless, with this measurement we explicitly demonstrate control of the hybridization between two fermionic levels with magnetic flux.

\begin{figure*}[t]
    \centering
    \includegraphics{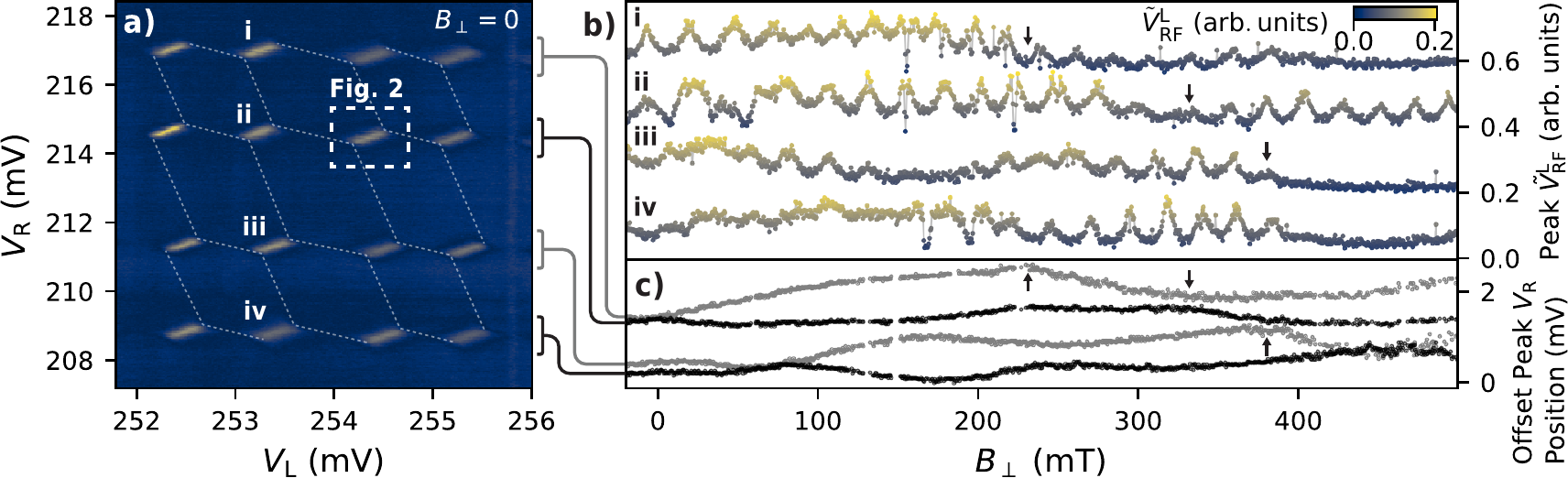}
    \caption{
      {\bf Flux-tunable hybridization across multiple dot levels.}
      \textbf{a)} CSD at zero applied field showing the window of 16 inter-dot transitions probed over a sweep of $B_\perp$.
      Dashed lines show the approximate boundaries of stable charge regions, while due to weak coupling of the QDs to the leads, only inter-dot transitions are visible in the resonator L signal.
      \textbf{b)} Peak signal deviation from Coulomb blockade $\vprfl$ of the four labeled transitions as a function of $B_\perp$, offset by \SI{0.18}{\arbunit}.
      \textbf{c)} Peak positions of inter-dot transitions in $\vpr$ coordinates relative to the lowest peak, averaged across all four columns of transitions shown in \textbf{a)}, and offset by \SI{2.32}{\milli\volt}.
      The offset voltages vary linearly with the addition energies of QDR, such that anticrossings in the positions correspond to anticrossings between electron states of QDR.
      Black arrows show points where correlation can be observed between the oscillation amplitude of $\vprfl$ and anticrossings of QDR states.
    }
    \label{fig:intermediateregime}
\end{figure*}

Next, in the same DQD regime, we study the variance of the oscillation amplitude in a broader field range and across multiple transitions, focusing on the 16 transitions shown in Fig.~\subref{fig:intermediateregime}{a}.
There, we plot the absolute deviation of the complex reflection signal of QDL's resonator from its average value in Coulomb blockade, denoted $\vprfl$.
The complex signal is a one-to-one function of the frequency shift of QDL's resonator \cite{Malinowski_2022}, and thus contains information about $\vert\teff\vert$.
An even-odd alternation in the transition spacing both along the $\vpl$ and $\vpr$ axes suggests that both QDs have spin degenerate levels with a finite level spacing in this window.
We sweep $B_\perp$, measuring new CSDs of the 16 transitions at a single measurement frequency adjusted to remain close to resonance.
From these CSDs, we extract the maximum $\vprfl$ signal and approximate peak position in gate space for all transitions.

We plot in Fig.~\subref{fig:intermediateregime}{b} the peak signal height for the column of transitions enumerated in Fig.~\subref{fig:intermediateregime}{a}, where for all four transitions, $h/e$-periodic oscillations of the peak height are clearly seen in some ranges of $B_\perp$.
There, we identify four distinct features.
First, regions in Fig.~\subref{fig:intermediateregime}{b} presenting a low signal and oscillation amplitude are visible, such as for transition iii at fields above \SI{400}{\milli\tesla}.
As is schematized in Fig.~\subref{fig:freqfits}{b-c}, this corresponds to large average $\vert\teff\vert$ and asymmetric barriers, where a smaller frequency shift and thus change in $\vprfl$ is expected.
Second, for smaller mean $\vert \teff\vert$ values the variation in signal with flux is much larger since $\vert\diff \Delta f_0^\mathrm{L}/\diff \vert\teff\vert\vert$ is larger, as seen for transition iv in the range \SIrange{280}{400}{\milli\tesla}.
Third, transition iv at low fields exhibits a substantial peak height indicating a small tunnel coupling but very weak oscillation contrast, suggesting that the tunnel barriers are tuned by $B_\perp$ to be substantially asymmetric in this field range.
Finally, for some transitions a sudden drop of the peak height to near zero appears near the oscillation maximum.
We expect this is a result of $\vert\teff\vert$ being small enough near the maximum peak height that thermal excitations and Landau-Zener transitions populate the excited DQD state, suppressing quantum capacitance (Suppl. Sec.~SVII) \cite{Ivakhnenko_2023,Gonzalez_Zalba_2016}.
Importantly, this also suggests that $t_\mathrm{B}\approx t_\mathrm{T}$ in those cases.

Differences between these scenarios are known to have consequences when sensing tunnel coupling to manipulate or measure qubits \cite{Petersson_2012,de_Jong_2019,Zheng_2019}.
Probing the tunnel coupling with gate sensing in the regime of very weak tunneling gives a sharp change in the resonator signal for small changes in $\vert\teff\vert$, allowing one to couple QDs weakly to the qubit of interest.
Conversely, the signal is then also sensitive to small changes in flux. 
Certain topological qubit proposals also rely on a substantial tunneling magnitude for their operation \cite{Leijnse_2012}.

To better understand Fig.~\subref{fig:intermediateregime}{b}, we now consider the influence of the specific electronic levels involved on tunnel coupling oscillation amplitude.
To this end, we plot the relative $\vpr$ position of inter-dot transitions averaged across all four columns in Fig.~\subref{fig:intermediateregime}{c}.
This position is proportional to the excitation energies of the different levels \cite{Stewart_1997,van_der_Wiel_2002}, and we observe that the QDR levels are nearly spin-degenerate at zero field.
Kinks can be seen in the peak positions, indicating (anti)crossings between levels of QDR.
At some fields, with examples highlighted by black arrows in Fig.~\subref{fig:intermediateregime}{b,c}, sudden changes in the average peak height and oscillation contrast of a transition appear correlated with anticrossings of QDR levels.
We hypothesize that variation with field of the wavefunction overlap of different levels, as well as the particular levels involved, can have a drastic effect on $t_\mathrm{T/B}$.
In particular, transitions between states of opposing spin have $\teff$ determined by spin-orbit coupling strength \cite{Hanson_2007,Nadj_Perge_2012,Han_2023}, while transitions between states of the same spin do not.
Given the large out-of-plane $g$-factor of these 2DEGs \cite{Moehle_2021}, it is impossible in this experiment to independently study spin and flux effects.
Contrarily, some changes in the mean peak height and oscillation contrast have no obvious correlation with QDR excitation energies, though we note that the QDL level involved also may have an effect.
Hence, for any application requiring hybridization readout between QD levels, the specific levels used must be optimized for a given magnetic field range.

\begin{figure}[t]
  \centering
  \includegraphics{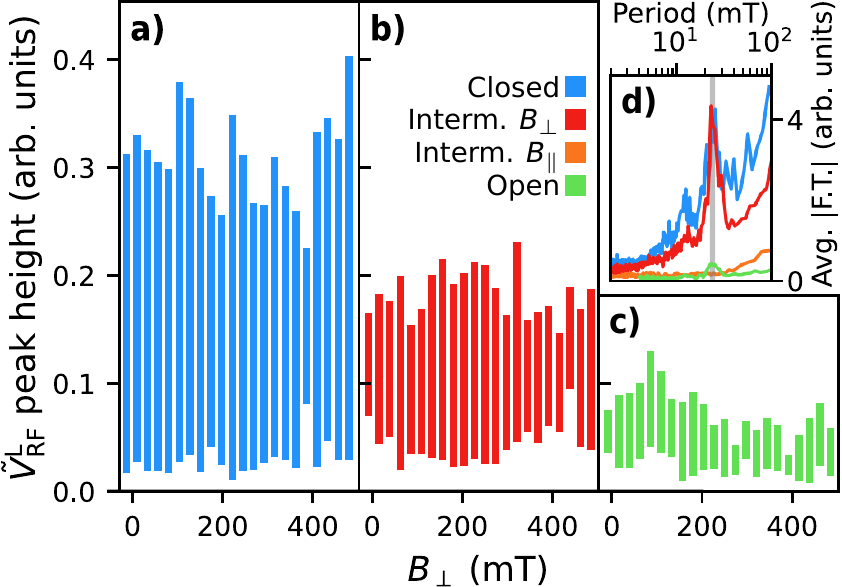}
  \caption{
    {\bf Contrast of Tunnel Coupling Variation in Different Regimes.}
    \textbf{a-c)} Bars showing maximal peak height variation on a single interdot transition spanning the distance between the smallest and largest observed $\tilde{V}_\mathrm{RF}^\mathrm{L}$ peak height, binned within one $h/e$ period of \SI{23.5}{\milli\tesla} and plotted for three different regimes of tunnel barrier tuning.
    Of the 16 inter-dot transitions tracked in each dataset, only the bar for the transition with the largest signal variation for each period is shown.
    \textbf{a)} summarizes a $B_\perp$ sweep in a regime of weak inter-dot tunneling with more negative barrier voltages, while \textbf{c)} shows data for strong tunneling and less negative barrier voltages.
    \textbf{b)} corresponds to the intermediate tunnel barrier data from Fig.~\ref{fig:intermediateregime}.
    The largest contrast in the signal generally occurs within the weak coupling regime.
    \textbf{e)} Absolute Fourier transforms in each regime averaged across all 16 transitions.
    Orange represents a sweep of the in-plane field for the same transitions and tuning as the intermediate regime.
    A vertical line shows the peak at \SI{23.5}{\milli\tesla}.
  }
  \label{fig:regimecomparison}
\end{figure}

Lastly, we compare the differences in tunnel coupling readout contrast for regimes of different $V_\mathrm{T/B}$ and thus average $t_\mathrm{T/B}$ values.
From Eq.~\ref{eq:teff} we expect that for nearly equal $t_\mathrm{B}$ and $t_\mathrm{T}$, large tunnel couplings should produce the best oscillation contrast, since the tunnel coupling ranges from $\vert t_\mathrm{T}\vert + \vert t_\mathrm{B}\vert$ to nearly zero.
Thusly, we conduct analogous measurements to those in the intermediate coupling regime of Fig.~\ref{fig:intermediateregime} for other coupling regimes, with results summarized in Fig.~\ref{fig:regimecomparison} and shown in more detail in Suppl. Sec.~SVIII.
Namely, we first bin the peak heights for a given regime into windows equal to the $h/e$ periodicity extracted from their average Fourier transform (Fig.~\subref{fig:regimecomparison}{d}).
Next, we plot bars spanning the minimum $\vprfl$ peak height to the maximum for whichever of the 16 transitions maximizes this difference in a given field bin.
In addition to the dataset from Fig.~\ref{fig:intermediateregime}, datasets for more negative (closed) and less negative (open) barrier gate voltages are shown in blue and green respectively.
As a control, we also show in orange data for an in-plane field sweep over the same transitions considered in Fig.~\ref{fig:intermediateregime}, where no oscillations are seen.
Compared to the red `intermediate' coupling regime, the more closed-off regime shows on average larger variation in peak height across a single $h/e$ period, due to the increased slope of $\Delta f_0^\mathrm{L}$ with flux as described above.
The open regime showed very weak oscillation contrast despite the tunnel barriers exhibiting similar resistances (Suppl. Sec.~SIII), suggesting that larger coupling regimes are more sensitive to slight asymmetries between $t_\mathrm{T}$ and $t_\mathrm{B}$.


Herein we have probed the tunability of hybridization between two electronic levels threaded by a magnetic flux.
Using gate-based RF reflectometry implemented in a phase-coherent InSbAs 2DEG, we measured $h/e$-periodic oscillations of tunnel coupling between the levels of two QDs arranged in a loop.
Even for nearly symmetrically tuned inter-dot tunnel barriers, the coupling was not generically suppressed at its minima, exhibiting a high degree of variability in magnitude and contrast of the tunnel coupling oscillations.
We inferred this variability to depend in part on the specific QD levels involved.
Finally, we found that, given the inherent difficulty of symmetrically tuning two tunnel barriers in parallel, the best contrast in signal for small changes in tunnel coupling was found to occur for relatively weak average inter-dot tunnel couplings \cite{de_Jong_2019}.
This work establishes a prerequisite for the readout of qubits formed in topological nanowires and Kitaev chains \cite{Plugge_2017,Karzig_2017,Sz_chenyi_2020,Liu_2022_ArXiv} and implements a new technique for the tuning of effective coupling between localized electronic states, concurrently illustrating its limitations.

\begin{acknowledgments}
    Raw data, analysis code, and scripts for plotting the figures in this publication are available from Zenodo \cite{zenodo}.
    
    C.G.P. and I.K. fabricated the device using a 2DEG heterostructure provided by D.X., C.T., and M.J.M..
    C.G.P. and M.C. conducted the measurements with input from L.H. and F.K.M..
    F.K.M. and S.G. supervised the project.
    C.G.P. analyzed the data and wrote the manuscript with input from all authors.

    The authors are grateful to J.V. Koski, L.P. Kouwenhoven, and F. Borsoi for helpful discussions and input on the manuscript, and to L.P. Kouwenhoven for initiating the project.
    The authors would also like to acknowledge financial support from Microsoft Quantum and the Dutch Research Council (NWO).
    F.K.M. acknowledges support from NWO under a Veni grant (VI.Veni.202.034).
\end{acknowledgments}

\bibliography{dqd_ring_bib}
\end{document}


\title{Supplementary information for ``Flux-Tunable Hybridization in a Double Quantum Dot Interferometer"}

\author{Christian G. Prosko}
\affiliation{QuTech and Kavli Institute of Nanoscience, Delft University of Technology, 2600 GA Delft, The Netherlands}

\author{Ivan Kulesh}
\affiliation{QuTech and Kavli Institute of Nanoscience, Delft University of Technology, 2600 GA Delft, The Netherlands}

\author{Michael Chan}
\affiliation{QuTech and Kavli Institute of Nanoscience, Delft University of Technology, 2600 GA Delft, The Netherlands}

\author{Lin Han}
\affiliation{QuTech and Kavli Institute of Nanoscience, Delft University of Technology, 2600 GA Delft, The Netherlands}

\author{Di Xiao}
\affiliation{Department of Physics and Astronomy, Purdue University, West Lafayette, Indiana 47907, USA}

\author{Candice Thomas}
\affiliation{Department of Physics and Astronomy, Purdue University, West Lafayette, Indiana 47907, USA}

\author{Michael J. Manfra}
\affiliation{Department of Physics and Astronomy, Purdue University, West Lafayette, Indiana 47907, USA}
\affiliation{School of Materials Engineering, Purdue University, West Lafayette, Indiana 47907, USA}
\affiliation{Elmore School of Electrical and Computer Engineering,
    Purdue University, West Lafayette, Indiana 47907, USA}

\author{Srijit Goswami}
\affiliation{QuTech and Kavli Institute of Nanoscience, Delft University of Technology, 2600 GA Delft, The Netherlands}

\author{Filip K. Malinowski}
\affiliation{QuTech and Kavli Institute of Nanoscience, Delft University of Technology, 2600 GA Delft, The Netherlands}

\date{\today}
\maketitle

\begin{figure}[h]
    \centering
    \includegraphics{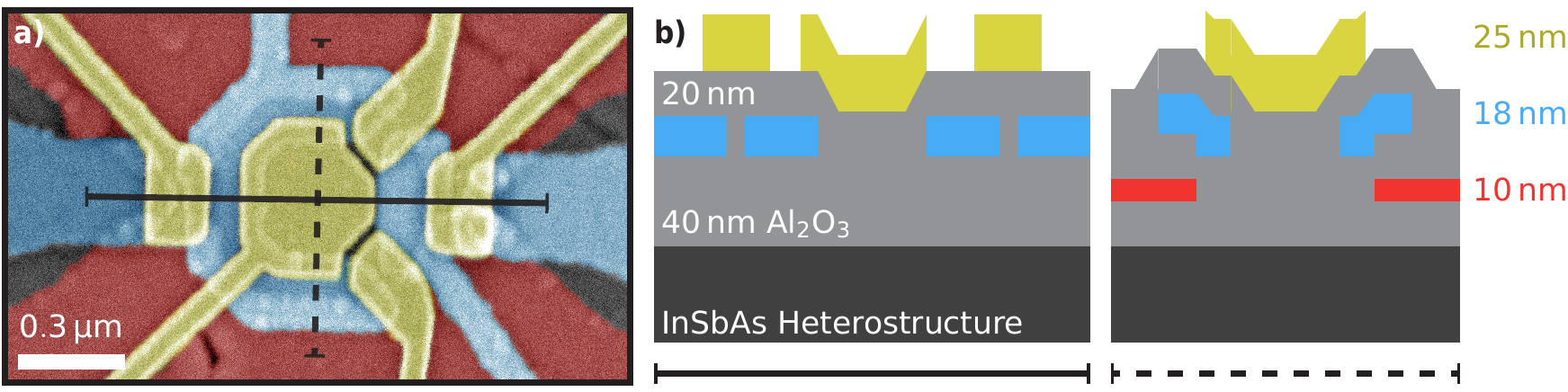}
    \caption{
        {\bf Device design and layer stack.}
        \textbf{a)} False-color scanning electron micrograph for a device nominally equivalent to the one measured on the same chip.
        The colors are encoded by gate layer, of which there are three, instead of by the gates' purpose as was done in Fig.~1a of the main text.
        \textbf{b)} Cross-sections approximately depicting the layer stack of the device along the solid and dashed lines shown in \textbf{(a)}.
        Thicknesses of the dielectric and $\mathrm{Ti}/\mathrm{Pd}$ gate layers are relatively to scale, but the widths are not, and the topography is only schematically depicted.
    }
    \label{fig:devicelayers}
\end{figure}

\section{Device Design \& Fabrication}
\label{sec:designandfab}

Here we describe in more detail the design considerations in fabricating the measured device.
One equivalent in design to the one measured from the same chip is shown in Fig.~\subref{fig:devicelayers}{a}.
Initially, the chip is covered with a \SI{<10}{\nano\meter} epitaxial layer of Al which was selectively etched away everywhere except in a region to the left and right of the pictured device to form leads, exposing the $\mathrm{InSb}_{0.86}\mathrm{As}_{0.14}$ two-dimensional electron gas (2DEG) heterostructure \cite{Moehle_2021}.
Next, the 2DEG was etched away except in a region close to the active device and along a roughly \SI{140}{\micro\meter} path connecting it to the Al leads, forming a mesa.
We then alternated between using atomic layer deposition to deposit roughly \SI{20}{\nano\meter} $\mathrm{Al}_2\mathrm{O}_3$ dielectric layers then evaporating Ti/Pd gate layers to form three electrically isolated gate layers.
Each layer also contains coarse gate leads (not shown), required to facilitate climbing the mesa.
The 2DEG mesa on which the device was fabricated conducts, so forming a loop required application of negative voltages both along the outer perimeter of the loop, as well as in the hole in the center.
Fabricating a double quantum dot (DQD) in this loop further necessitated plunger gates to tune the chemical potential of the quantum dots (QDs) and gates to form barriers between them and to the contacts.
One option to satisfy these requirements is to pattern depletion gates in a layer above the plunger gates needed to tune the QDs, however in this case the leads of the lower layer gates were found in previous devices to screen the depletion gate voltage and prevent forming a stable loop.
Hence, it was topologically required to fabricate three gate layers in order to both have an outer depletion gate underneath the plunger and barrier gates, as well as a central depletion gate which can cross over the plunger gates to deplete the center of the loop.
The corresponding layer stack is schematized in Fig.~\subref{fig:devicelayers}{b}.
A third gate layer had the added advantage that tunnel barriers could be made effectively more narrow, since barrier gates in the third layer may overlap with plunger gates in the second layer.

\newpage
\section{Quantum Ring Measurements}
\label{sec:quantumring}

\begin{figure}[h]
    \centering
    \includegraphics{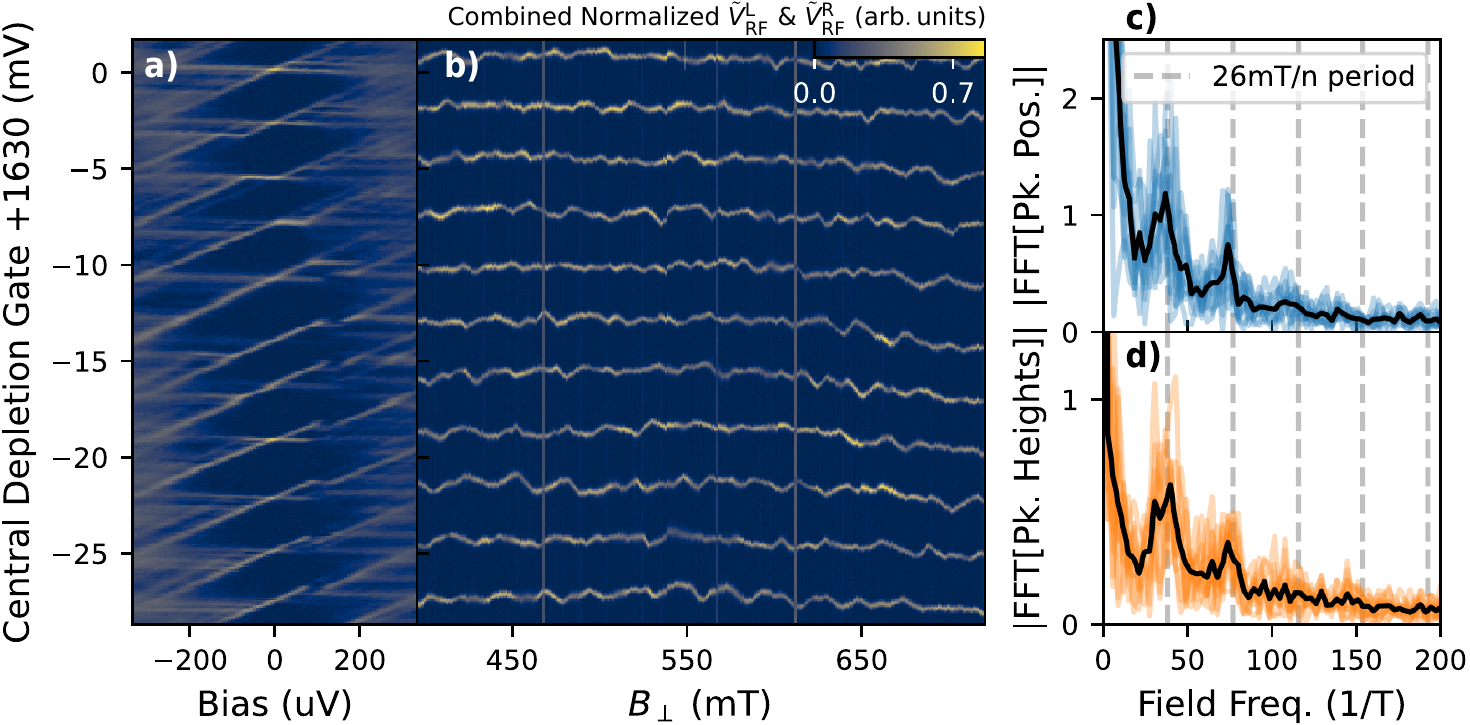}
    \caption{
        {\bf Field Dependence of Coulomb resonances in the ring-shaped QD regime.} \textbf{a)} Coulomb diamonds of the loop-shaped dot formed under the $\vpl$ and $\vpr$ electrodes measured with gate reflectometry at $B_\perp=\SI{950}{\milli\tesla}$ out-of-plane field.
        Its chemical potential is tuned by the central depletion gate shown in Fig.~\subref{fig:devicelayers}{a}, and it is formed by lowering $\vbs$ and $\vbd$ voltages to form tunnel barriers to the leads.
        Throughout these diamonds, an orbital spacing of at least \SI{45}{\micro\electronvolt} is observed.
        \textbf{b)} Zero-bias Coulomb resonances measured as a function of $B_\perp$.
        The signal is constructed from the multiplexed signals of resonators L and R by centering each complex signal around its Coulomb blockade value, normalizing it, then taking the absolute average.
        Irregular oscillations appear in all resonances \cite{Fuhrer_2001}.
        Absolute Fourier transforms of the peak voltage position \textbf{(c)} and the peak signal \textbf{(d)}.
        Colored lines show Fourier transforms of individual Coulomb resonances, while the black lines show their mean. Gray lines highlight frequencies corresponding to an $h/e$ flux periodicity of \SI{26}{\milli\tesla} as well as higher harmonics.
    }
    \label{fig:quantumring}
\end{figure}

\newpage

\section{Tuning Symmetric Parallel Tunnel Barriers}
\label{sec:tunnelbarriers}

\begin{figure}[!h]
    \centering
    \includegraphics{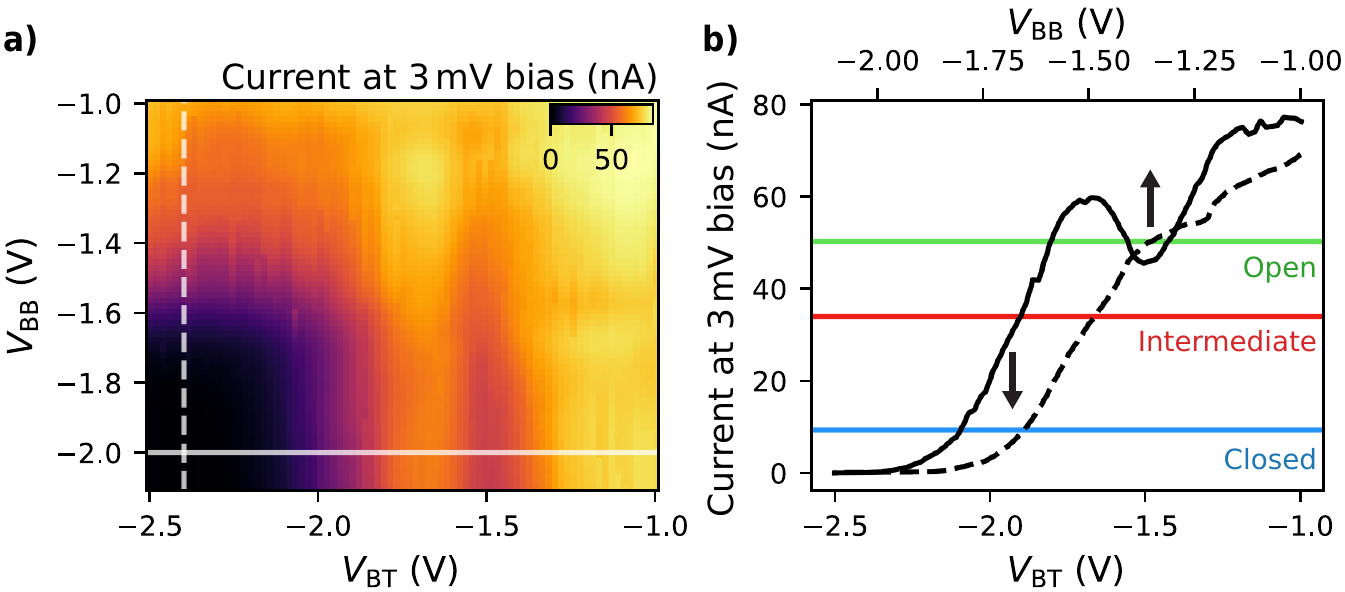}
    \caption{
        {\bf Pinch-off scans for approximately symmetric barrier tuning.} \textbf{a)} Current through the device at \SI{3}{\milli\volt} applied bias voltage as a function of $\vbt$ and $\vbb$, tuned into an otherwise open loop.
        The roughly rectangular shape of the zero-current region implies a weak cross-coupling between gates BT and BB.
        Linecuts where BT or BB are closed (white lines) can thus be used to select barrier voltages for roughly equal resistance.
        \textbf{b)} Linecuts from the current map of \textbf{(a)}.
        To tune for the intermediate coupling regime of Fig.~3 in the main text (red), or the more closed off (blue) and open (green) regimes described in Fig.~4, $\vbt$ and $\vbb$ voltages are chosen such that when the opposite barrier is pinched off, they both admit roughly the same current.
        The relatively large bias reduces the influence of QD states under the barriers on the measurement.
    }
    \label{fig:barriertuning}
\end{figure}

\section{Coulomb Diamonds}
\label{sec:coulombdiamonds}

\begin{figure}[!h]
    \centering
    \includegraphics{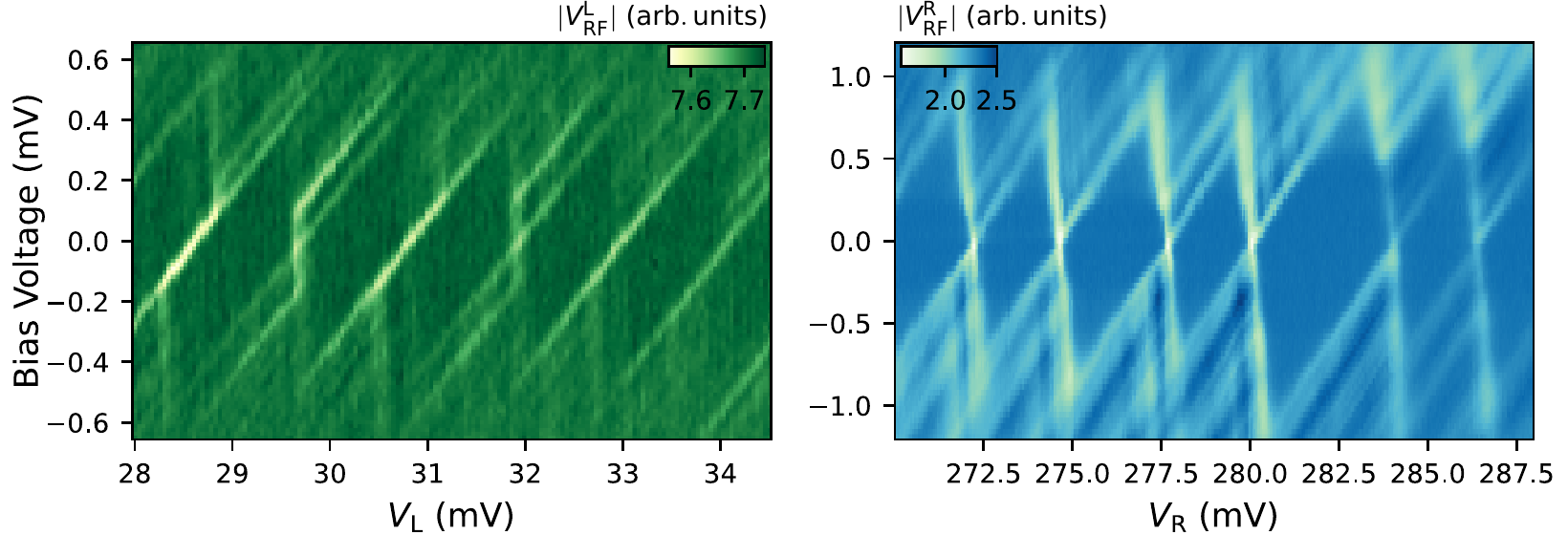}
    \caption{
        {\bf Coulomb diamonds of QDL (left) and QDR (right).}
        The single QDs are tuned such that both the relevant lead barrier as well as $\vbt$ and $\vbb$ are in a weak tunneling regime.
        Magnitude of the reflectometry signal near the resonance frequency of their respective plunger gates' resonators is plotted.
        A varying but finite level energy spacing is observed for both QDs larger than the linewidth.
    }
    \label{fig:coulombdiamonds}
\end{figure}

\section{Fitting Procedure for Extraction of Tunnel Couplings}
\label{sec:fitprocedure}

Herein we detail the procedure used to extract the effective tunnel coupling magnitude of a DQD ($\vert \teff\vert$ in the main text), given a charge stability diagram (CSD) spanning an inter-dot charge transition with a frequency-dependent response measured at each point for a resonator coupled to one of the QD's gates.
The parametric capacitance for a gate at voltage $V_\mathrm{g}$ primarily coupled to a single charge island or QD (indexed by $i$) out of multiple potentially coupled islands is
\begin{equation}
    C_\mathrm{p} = \tilde{\alpha}_i\vert e\vert \frac{\diff \braket{\hat{n}_i}}{\diff V_\mathrm{g}},
\end{equation}
where $\braket{\hat{n}_i}$ is the expectation value of charge on QD $i$ and $\tilde{\alpha}_i$ is a lever arm of the gate's coupling to the quantum modified by mutual capacitances of this QD to other charge islands in the system, see Sec.~\ref{sec:paracap} for further details.
In essence, the large inter-dot capacitance of the system when tuned into the DQD regime (as can be inferred from the inter-dot transition width in gate space relative to the spacing between transitions in Fig.~3a \cite{van_der_Wiel_2002}) lowers the effective lever arm of the gate to the sensed QD.
Consequently, we must fit for $\tilde{\alpha}_i$ independently, since it is not expected to agree with the lever arms extractable from the Coulomb diamond measurements of Fig.~\ref{fig:coulombdiamonds}.
This parametric capacitance can be calculated from the fitted resonator frequency $f_0$ as $C_\mathrm{p} = 1/4\pi^2Lf_0^2 - C$ where $L$ and $C$ are the resonator's bare inductance and capacitance, respectively.
In practice, we approximate $L$ at zero magnetic field as its simulated value for the resonator's inductor coil.
We calculate $C$ from the resonance frequency in Coulomb blockade, where $C_\mathrm{p}$ is assumed zero.
At each value of the out-of-plane magnetic field $B_\perp$, we assume that in Coulomb blockade the only shift in the resonator frequency is due to changes in $L$, such that from frequency fits at each field we can extract $L(B_\perp)$ assuming $C(B_\perp)$ is fixed.
Thus, the parameters $L$ and $C$ are fixed by measurements and not varied in the subsequent fits described below.

As an explicit model for parametric capacitance, we consider the model of Refs.~\cite{Mizuta_2017,Esterli_2019} for a DQD coupled to a phonon bath.
Near an inter-dot transition, this model considers two charge states with an excess electron residing either on a discrete fermionic mode of the sensed QD, or a mode of a second QD.
These two modes are coupled by tunnel coupling $\teff$, and the detuning between their energies is given by $\varepsilon = \tilde{\alpha}_i(V_\mathrm{g}-V_\mathrm{g}^\mathrm{off})$ where the offset $V_\mathrm{g}^\mathrm{off}$ determines the transition position in gate space.
In this model, the parametric capacitance is found to be
\begin{equation}
    C_\mathrm{p}
    =
    \underbrace{\frac{(e\tilde{\alpha}_i)^2}{8}\frac{\vert\teff\vert^2}{(\Delta E)^3}\tanh{\left(\frac{\Delta E}{2k_BT}\right)}}_{\equiv C_\mathrm{q}(\varepsilon)}
    + \frac{(e\tilde{\alpha}_i)^2}{4k_BT}\left(\frac{\varepsilon}{\Delta E}\right)^2\frac{\gamma^2}{\omega^2+\gamma^2}\cosh^{-2}{\left(\frac{\Delta E}{2k_BT}\right)},
\end{equation}
where $\Delta E \equiv \sqrt{(\varepsilon/2)^2+\vert\teff\vert^2}$ is the energy splitting of the charge qubit and $\omega$ is the angular resonator measurement frequency.
The first term above corresponds to quantum capacitance while the second corresponds to so-called tunneling capacitance.
The parameter $\gamma$ quantifies incoherent tunneling due to phonon absorption and emission, and in principle is another parameter we must include in our fit of $C_\mathrm{p}$ to extract $\vert\teff\vert$.

\begin{figure}[t]
    \centering
    \includegraphics{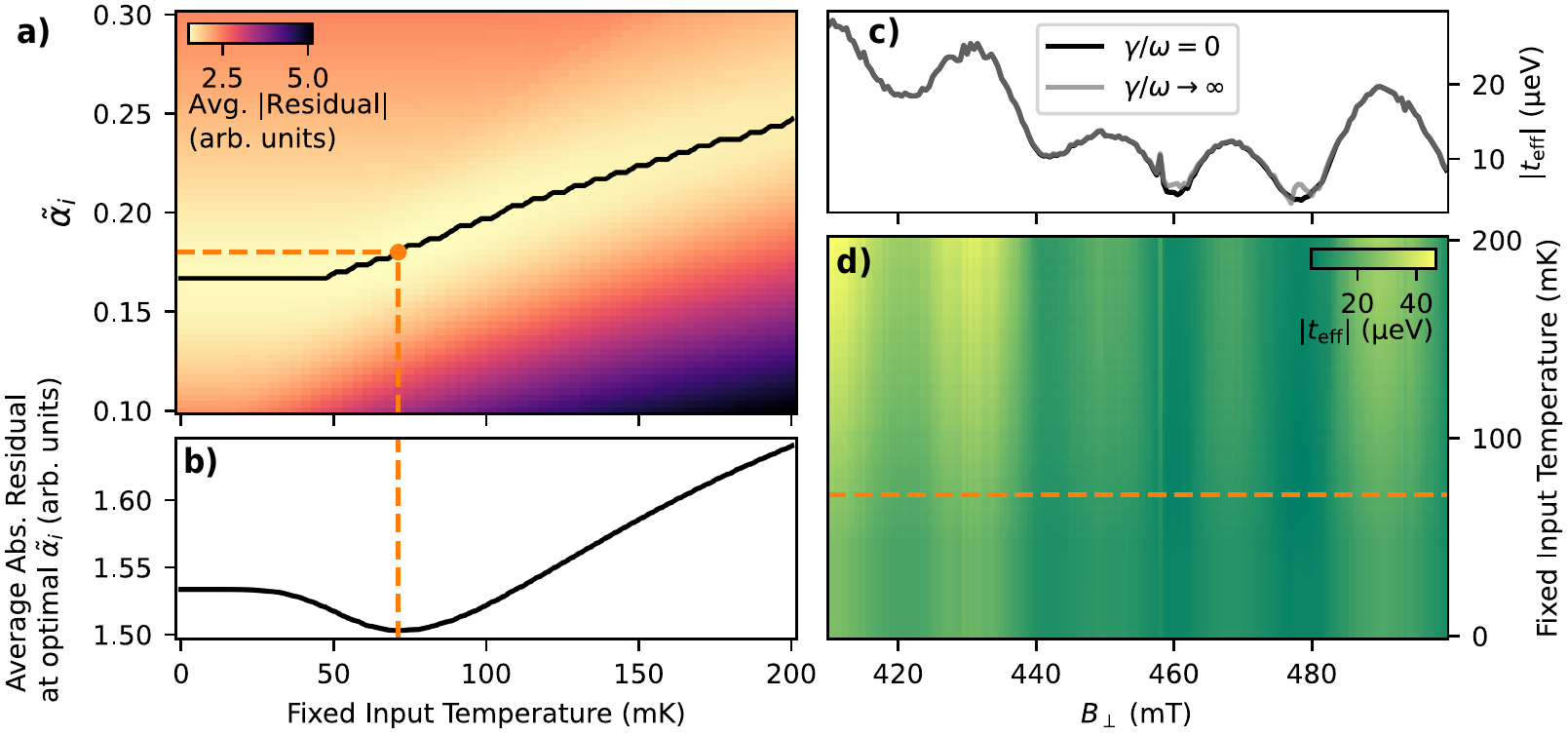}
    \caption{
        {\bf Optimization of tunnel coupling fits.}
        \textbf{a)} The mean absolute residual difference between the fit $C_\mathrm{p}$ lineshape of the inter-dot charge transition as a function of $V_\mathrm{L}$ defined in the main text and the $C_\mathrm{p}$ values extracted from fit frequency shifts of QDL's resonator.
        The black line shows the fixed $\tilde{\alpha}_i$ value minimizing the residual error for each fixed $T$.
        \textbf{b)} The mean residual error with $\tilde{\alpha}_i$ fixed at its optimal value shown in \textbf{a)} for each fixed value of $T$.
        A clear minimum is found at $T=\SI{71}{\milli\kelvin}$ and $\tilde{\alpha}_i=0.18$.
        \textbf{c)} The extracted $\vert t_\mathrm{eff}\vert$ for zero tunneling capacitance ($\gamma=0$) and maximal tunneling capacitance, which saturates as $\gamma\rightarrow\infty$.
        The presence of tunneling capacitance has a negligible effect on $\vert t_\mathrm{eff}\vert$ except at very small $\vert t_\mathrm{eff}\vert$.
        \textbf{d)} Fit $\vert t_\mathrm{eff}\vert$ with $\tilde{\alpha}_i$ fixed to the value minimizing fit error for each value of fixed temperature.
    }
    \label{fig:fitoptimization}
\end{figure}

A resistive contribution to the effective impedance of the sample known as Sisyphus conductance arises, however, whenever there is substantial tunneling capacitance \cite{Mizuta_2017,Esterli_2019}, which would lower the resonator internal quality factor near the transition.
In our fits of the frequency-dependent CSDs, the change in resonator quality factor was not discernible at the inter-dot transition, indicating that Sisyphus resistance and likely tunneling capacitance can be neglected in our fits.
This also indicates that all information about inter-dot tunneling is contained in the frequency shift $\Delta f_0$, such that we may solely fit $\Delta f_0(V_\mathrm{g})$ to extract $\vert\teff\vert$, rather than simultaneously fitting the frequency shift and quality factor.
Regardless, in Fig.~\subref{fig:fitoptimization}{c} we show that maximizing the contribution of tunneling capacitance leads to a negligible change to the extracted $\vert t_\mathrm{eff}\vert$ except for very small tunnel couplings.
Hence, we neglect tunneling capacitance for the fits of Fig.~2d.

Under these constraints, we extract a fitted $C_\mathrm{p}(V_\mathrm{g})$ from fitted $\Delta f_0$ and our knowledge of $L$ and $C$ described above, and fit the result to
\begin{equation}
    C_\mathrm{p}
    = C_\mathrm{q}(\tilde{\alpha}_i (V_\mathrm{g}-V_\mathrm{g}^\mathrm{off})) + C_\mathrm{off}
\end{equation}
with $C_\mathrm{q}$ as defined above.
In fact, we select five rows of the gate voltage near the center of the transition and fit them simultaneously with the same $\vert t_\mathrm{eff}\vert$, $\tilde{\alpha}_i$, and $T$, but allow for a different $C_\mathrm{off}$ and $V_\mathrm{g}^\mathrm{off}$ for each row.
In other words, we fit multiple traces for values of the other QD's gate voltage near the center of the charge transition in the charge stability diagram.
The offset $C_\mathrm{off}$ accounts for errors in converting from $\Delta f_0$ to $C_\mathrm{p}$.
These parameters are fitted independently for each row.

Since $T$ and $\tilde{\alpha}_i$ should be roughly the same at all fields, we sweep different fixed values of these parameters iteratively and choose the values which lead to a minimum total residual across all magnetic field values.
We found a global optimum of $T=\SI{71}{\milli\kelvin}$ and $\tilde{\alpha}_i=0.18$ which minimized the mean absolute fit residual error, see Fig.~\subref{fig:fitoptimization}{a,b}.
This temperature is larger than the roughly \SI{20}{\milli\kelvin} temperature of the dilution refrigerator used, which is not unexpected since electron temperature may be raised by connection to higher temperature cables and electronics \cite{Kouwenhoven_1997}.
Lastly, in Fig.~\subref{fig:fitoptimization}{d}, we observe that the oscillation amplitude of $\vert t_\mathrm{eff}\vert$ does vary with increasing temperature used in the fits (with $\tilde{\alpha}_i$ fixed at the optimum shown in Fig.~\subref{fig:fitoptimization}{a}), but the oscillations of $\vert t_\mathrm{eff}\vert$ are consistently present with a period of one flux quantum.

\section{Capacitance Formula Including Mutual Capacitances}
\label{sec:paracap}

In order to determine the degree to which mutual capacitances between QDs suppress parametric capacitance, we follow the approach of Refs.~\cite{Mizuta_2017,Esterli_2019} to derive an expression for parametric capacitance, additionally considering mutual capacitance effects to second order.
We consider the case of $N$ charge islands coupled capacitively to a single gate voltage $V_{\mathrm{g}}$ via capacitances $C_{\mathrm{g}i}$ for $i\in\{1,2,...,N\}$, with mutual capacitances between the islands of $C_{ij}$ for $i\neq j$, and other capacitive couplings to ground encompassed by an environmental capacitance $C_{\mathrm{e}i}$.
The latter includes any capacitances to lead reservoirs, for example.
We refer to the total capacitance of each island as $C_i \equiv C_{\mathrm{g}i} + C_{\mathrm{e}i} + \sum_{j\neq i}C_{ij}$.
Note that by definition, we have $C_{ij} = C_{ji}$.
The total differential capacitance $C_{\mathrm{diff}}$ as seen by $V_{\mathrm{g}}$ can then be written as the sum over differential capacitance contributions of each island
\begin{equation}
    C_{\mathrm{diff}}
    = \sum_{i=1}^{N}\frac{\diff \braket{Q_i}}{\diff V_{\mathrm{g}}}
    = \frac{\diff\sum_{i=1}^{N}\braket{Q_i}}{\diff V_{\mathrm{g}}}
\end{equation}
where $Q_i$ is the total effective charge on the capacitor $C_{\mathrm{g}i}$ as seen by $V_{\mathrm{g}}$ and the angular brackets denote the statistical average of the charge.
In general, this average must include thermodynamic, quantum mechanical, and driving effects.

To solve this expression, we write $\braket{Q_i}$ in terms of known capacitances and the expectation values $\braket{\hat{n}_i}$ of electron number on each island with charge number operator $\hat{n}_i$.
First, by definition of the gate capacitances we may write $\braket{Q_i} = C_{\mathrm{g}i}(V_\mathrm{g}-V_i)$ where $V_i$ is the electrostatic potential on island $i$.
On average, we can write the charge expectation value on island $i$ as a sum over all of the voltage induced charges from each capacitor
\begin{equation}
    -\vert e\vert\braket{\hat{n}_i}
    = C_{\mathrm{g}i}(V_i - V_{\mathrm{g}}) + \sum_{j\neq i} C_{ij}(V_i-V_j) + C_{\mathrm{e}i}V_i
\end{equation}
with $e$ being the electron charge \cite{van_der_Wiel_2002}.
Solving for $V_i$ and recalling the definition of $C_i$, we find
\begin{equation}
    V_i
    = \frac{1}{C_i}\left(C_{\mathrm{g}i}V_\mathrm{g}+\sum_{j\neq i}C_{ij}V_j - \vert e\vert \braket{\hat{n}_i}\right).
\end{equation}
By substituting this result for each $V_j$ into the original expression for $V_i$, we may recursively generate expressions for $V_i$ to higher and higher orders in the mutual capacitance lever arms $C_{ij}/C_i$.
Doing so twice, substituting the result into the definition of $\braket{Q_i}$, and using the resulting expression to calculate $C_{\mathrm{diff}}$, we find
\begin{equation}
    C_{\mathrm{diff}}
    = C_\mathrm{geom} + C_\mathrm{p} + \mathcal{O}(C_{ij}^3/C_i^3)
\end{equation}
with contributions from a constant geometric capacitance
\begin{equation}
    C_\mathrm{geom}
    \equiv \sum_{i=1}^N\alpha_i\left[C_i-C_{\mathrm{g}i} - \sum_{j\neq i}\left(\alpha_jC_{ij} + \sum_{k\neq j}\frac{C_{ij}C_{jk}}{C_j}\alpha_k\right)\right]
\end{equation}
and a $\braket{\hat{n}_i}$-dependent parametric capacitance:
\begin{equation}
    C_\mathrm{p}
    \equiv \sum_{i=1}^N\left[
        \alpha_i
        + \sum_{j\neq i}\left(
        \alpha_j\frac{C_{ij}}{C_j}
        +\sum_{k\neq j}\alpha_k\frac{C_{ij}C_{jk}}{C_iC_k}
        \right)\right]
    \vert e\vert \frac{\diff\braket{\hat{n}_i}}{\diff V_{\mathrm{g}}}
\end{equation}
where we have defined the bare lever arms $\alpha_i \equiv C_{\mathrm{g}i}/C_i$.

Hence, in addition to large mutual capacitances renormalizing a coupled island's lever arm by increasing $C_i$, there is an additional renormalization factor due to mutual capacitances increasing the effective lever arm. The lowest-order of the latter corrections are multiplied by the cross-capacitive lever arms $\alpha_j\ll 1$, however.
Note additionally that as $V_\mathrm{g}$ tunes the islands near an inter-dot charge transition between islands $i$ and $j$, the transfer of an electron by this tuning implies $\diff\braket{\hat{n}_i}/\diff V_{\mathrm{g}} \approx -\diff\braket{\hat{n}_j}/\diff V_{\mathrm{g}}$ so that cross-capacitances $C_{\mathrm{g}j}$ between the gate voltage and islands other than the island it is designed to sense suppresses the parametric capacitance signal at these transitions \cite{Mizuta_2017,Esterli_2019}.
From the slope of successive triple points across multiple inter-dot transitions, these cross capacitances are estimated to be negligible in the measured regimes of this experiment.
In this limit, where $V_{\mathrm{g}}$ primarily couples to a single island $i$, but the island itself has relatively larger mutual capacitances to the other islands, we discard terms of the order $C_{ij}\alpha_j/C_j$ for $j\neq i$ but preserve terms to second order in $C_{ij}/C_j$ when multiplied by $\alpha_i \gg \alpha_j$, leading to
\begin{equation}
    C_\mathrm{p}
    \sim \left(1 + \sum_{j\neq i}\frac{C_{ij}^2}{C_i^2}\right)\alpha_i\vert e\vert\frac{\diff\braket{\hat{n}_i}}{\diff V_{\mathrm{g}}}
    = \frac{1 + \sum_{j\neq i}C_{ij}^2/C_i^2}{C_{\mathrm{e}i}+C_{\mathrm{g}i} + \sum_{j\neq i}C_{ij}}C_{\mathrm{g}i}\vert e\vert \frac{\diff\braket{\hat{n}_i}}{\diff V_{\mathrm{g}}}
    \hspace{10mm}
    \begin{array}{ll}
        C_{ij}/C_i, \alpha_j\ll 1 \\
        \alpha_j\ll C_{ij}/C_i
    \end{array}
    \hspace{5mm}
    \forall j\neq i.
\end{equation}

\section{Quantum Capacitance Supppression due to Landau-Zener Transitions}
\label{sec:LZTs}

Landau-Zener transitions (LZTs) make the used capacitance model inapplicable for small values of $\vert\teff\vert\lesssim\sqrt{\hbar \alpha\delta \vrf f_0}$, where $\delta \vrf$ is the resonator's oscillating voltage amplitude, $\alpha$ is its lever arm to the QD, and $f_0$ is the resonator frequency \cite{Ivakhnenko_2023}.
There LZTs become frequent, biasing the system towards equal occupation of the excited and ground charge states where quantum capacitance is zero \cite{de_Jong_2019}.
For a DQD with a short decoherence time, and at zero detuning from the charge transition, the probability of a LZT occurring twice in a resonator cycle is $e^{-2\vert\teff\vert^2/\hbar\alpha\delta \vrf f_0}$ \cite{Ivakhnenko_2023,Gonzalez_Zalba_2016}.
Due to the sinusoidal nature of the oscillating voltage, a LZT occurring twice in a cycle means that the tunneling electron spends an equal amount of time in the excited DQD state as in the ground state.
In other words, the population of the excited state is equal to the population of the ground state when this probability is one.
Hence, we expect quantum capacitance to be eventually suppressed for small enough $\vert\teff\vert$, since LZTs become more probable as $\vert\teff\vert$ becomes smaller for fixed $\delta \vrf$.
Thermal redistribution also becomes important for small $\vert\teff\vert$, further suppressing the frequency shift \cite{Mizuta_2017,Esterli_2019}.

\section{Field-Dependence of Peak Heights in Different Coupling Regimes}
\label{sec:fullpeakdata}

\begin{figure}[t]
    \centering
    \includegraphics{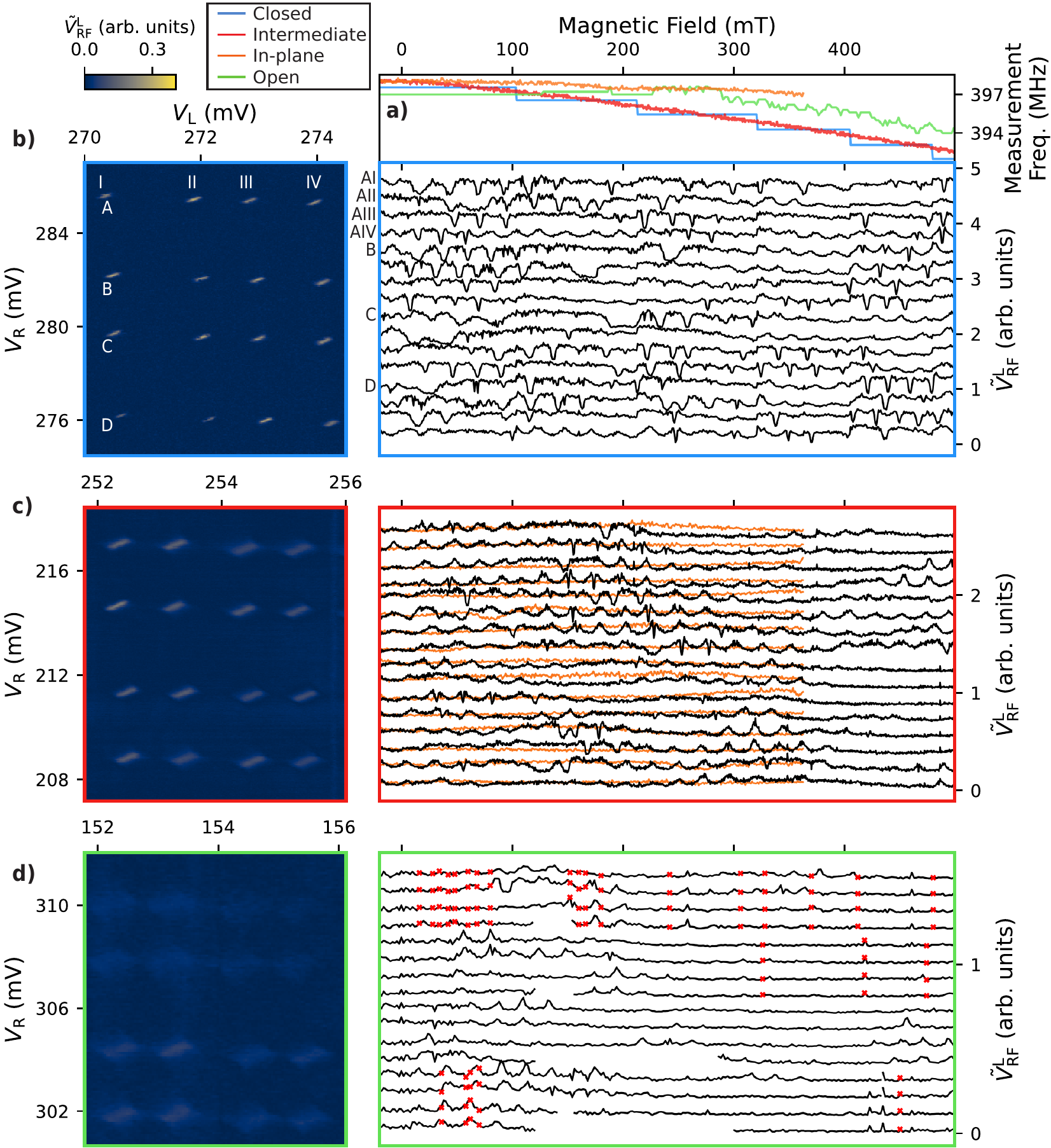}
    \caption{
        {\bf Field-dependence of inter-dot charge transitions in different regimes of tunnel coupling.}
        \textbf{a)} Measurement frequency for resonator L used at each out-of-plane field value $B_\perp$ for the three different regimes of tunneling strength investigated as well as for an in-plane field $B_\parallel$ sweep for the same transitions of the intermediate regime.
        \textbf{b-d)} Field-dependence data for the closed \textbf{(b)}, intermediate \textbf{(c)}, and open \textbf{(d)} tunnel coupling regimes.
        These correspond to voltages $(V_\mathrm{BT},V_\mathrm{BB})=(-2.1,-1.65)\si{\volt}$, $(-1.9,-1.49)\si{\volt}$, $(-1.82, -1.34)\si{\volt}$, for the closed, intermediate, and open regimes respectively.
        $\vbs$ and $\vbd$ were tuned to a very weak tunneling regime of $\vbs=\SI{-2.05}{\volt}$ and $\vbd=\SI{-2.75}{\volt}$, except in the closed regime where $\vbs=\SI{-2.5}{\volt}$.
        \emph{Left:} CSDs measured at zero magnetic field, plotting the reflected signal magnitude $\tilde{V}_\mathrm{RF}^\mathrm{L}$ from resonator L centered about the Coulomb blockade value.
        \emph{Right:} Field dependence of the peak deviation from Coulomb blockade for the 16 inter-dot transitions shown in the CSDs, offset by 0.3 \textbf{(b)}, 0.17 \textbf{(c)}, and \SI{0.09}{\arbunit} \textbf{(d)} for clarity.
        Peak heights in \textbf{c)} for the $B_\parallel$ sweep are plotted in orange.
        In \textbf{d)}, a stray resonance appeared which occluded inter-dot transitions for some transitions in a wide window.
        This resonance interfered with extraction of the peak signal height, and so appears as a gap in the plot.
        Red markers denote points at which charge jumps appeared in the search window used to extract the peak signal height.
    }
    \label{fig:fullpeakdata}
\end{figure}

In this section the full datasets from which Fig.~4 of the main text was constructed are shown in Fig.~\ref{fig:fullpeakdata}, including the dataset used in Fig.~3 of the main text.
The four datasets are measured in three different regimes of inter-dot barrier gate voltage strengths, denoted the `closed', `intermediate', and `open' regimes ordered from the strongest to the weakest barrier gate voltages separating QDL and QDR.
Though not shown in the figure, in the closed regime at fixed field values, some transitions occasionally exhibited a jitter from row to row in $V_\mathrm{L}$-space.
This may be due to very weak coupling from the DQD to the leads resulting in electrons tunneling on to the DQD stochastically as the gate is swept, and may result in unphysical additional suppression of the peak height for some fields.
Nonetheless, the prominent peak of the Fourier transform of this data at a periodicity of one flux quantum (shown in Fig.~4 of the main text) indicates that the sharp dips in the data truly correspond to a suppression of the signal periodically as a function of flux.

\bibliography{dqd_ring_bib}